\documentclass[nonacm,screen]{acmart}

\usepackage{array}
\usepackage{tabularx}
\usepackage{float} 

\begin{document}

\author{Woohyeuk Lee}
\affiliation{%
  \institution{University of Texas at Austin}
  \city{Austin}
  \country{USA}
}
\email{woohyeuk.lee@utexas.edu}

\author{Hanlin Li}
\affiliation{%
  \institution{University of Texas at Austin}
  \city{Austin}
  \country{USA}
}
\email{lihanlin@utexas.edu}

\author{David Gray Widder}
\affiliation{%
  \institution{University of Texas at Austin}
  \city{Austin}
  \country{USA}
}
\email{david.widder@austin.utexas.edu}

\renewcommand{\shortauthors}{Lee et al.}

\title{Open at the Edge, Captured at the Center: llama.cpp and the Political Economy of Local AI Inference}

\begin{abstract}
Critical scholarship on open AI has focused on model releases and cloud ecosystems, leaving the local inference infrastructure that makes open-weight models runnable on user-owned devices largely unexamined. We address this gap through a mixed-methods analysis of \texttt{llama.cpp}, combining 7,681 merged pull requests from March 2023 through March 2026 with repository discussions, corporate statements, and contributor blogs. We show that local inference broadens participation at execution while relocating capture into the infrastructure that makes execution possible. Through hardware backends, model integration labor, and Hugging Face's February 2026 absorption of the project, we document how control shifts to hardware vendors, model distributors, and core maintainers while model owners and individual contributors bear the cost of making models runnable. These dynamics suggest that preserving openness outside the cloud requires attention to the infrastructure that makes models runnable, not just to the models themselves. This calls for policy mechanisms---analysis of format dependencies and vendor influence, model compatibility requirements, and sustained public funding for inference tooling---that extend beyond model release conditions to the infrastructure layer.
\end{abstract}



\keywords{open source software, large language models, local inference, llama.cpp, model hubs, open-weight models, political economy, AI governance}

\maketitle

\section{Introduction}

Large language models reach users predominantly through cloud APIs \cite{nagleLatentRoleOpen2025}. In that arrangement, access to AI is not only access to a model. It is also access through provider-owned compute, provider-curated model catalogs, per-token billing, uptime guarantees, and rate limits \cite{vipra2023computational, grahamDigitalWorkPlanetary2022, luitse2024platform}. These layers shape what can be built, by whom, and under what terms. Even when a model is technically available, the practical conditions of using it are organized by the infrastructure through which it is served. That concentration of infrastructure has motivated alternatives that shift computation outward---but moving the site of execution does not automatically displace the sites of power. \texttt{llama.cpp} illustrates this dynamic: it concentrates the hardware, model-distribution, and maintenance dynamics that make local inference practical. Founded in March 2023 as a CPU-first inference tool for Meta's LLaMA weights, it was explicitly framed against cloud dependence\footnote{The ``manifesto'' of \texttt{llama.cpp}: \url{https://github.com/ggml-org/llama.cpp/discussions/205}} before expanding into a heterogeneous platform supporting many model families, device-specific execution paths, and downstream tools. In February 2026, its founding team joined Hugging Face\footnote{\url{https://github.com/ggml-org/llama.cpp/discussions/19759}}. A project that began as an alternative to cloud-mediated AI now has its core maintenance labor housed inside a model distributor whose business model depends on remaining the default hub for open-model workflows.

The analysis is organized around three questions. \textbf{First}, who participates in the development of local inference infrastructure, and what incentives drive their participation? \textbf{Second}, how does the political economy of this layer differ from cloud-centered accounts that currently dominate critical scholarship on open AI? \textbf{Third}, how did Hugging Face come to occupy a dominant position in local inference infrastructure, and what does its trajectory reveal about the structural conditions under which community-governed AI infrastructure persists or erodes? To address these questions, we draw from a corpus of 7,681 merged pull requests to \texttt{ggml-org/llama.cpp} through March 2026, extract the paths of files they change to identify what contribution they make, and link them to their authors' affiliations (or non-affiliations for individual contributors). We pair the PRs with corporate public statements, contributor blogs, and repository discussions that let us read quantitative patterns against stated incentives.

Our central contribution is to extend critiques of open-washing and corporate capture from model artifacts and cloud compute to local inference infrastructure, and present that openness at the artifact does not prevent enclosure at the infrastructure---a dynamic we track through how hardware vendors, model distributors, and upstream maintainers shape what it means for a model to be truly runnable. Where Widder et al. find that the capacity to train foundation models is concentrated in the same firms that rent the compute to run them \cite{widderOpenBusinessBig2023, widderWhyOpenAI2024}, we document a parallel concentration at the point where open models meet user devices. This concentration is mediated through hardware vendor strategies that turn compatibility work into footholds and steer ostensibly open device-support paths toward proprietary extensions, and through a distributor whose position across model distribution, model-file conversion, and inference-tool download and loading defaults approaches what Callon calls an obligatory passage point \cite{callonElementsSociologyTranslation1984}. Local inference is thus neither a solved alternative to the cloud nor structurally immune to the dynamics that have produced concentration elsewhere in AI.

\section{Related Works}

\subsection{Phenomenon of open models}

Scholarship on open models first centered the model artifact itself, asking which components of an AI release are actually open and what forms of access, control, and risk different release strategies create. Solaiman's gradients of openness \cite{solaimanGradientGenerativeAI2023} and component-level frameworks by Liesenfeld et al., White et al., and Bommasani et al. show that weights, code, data, documentation, licenses, and evaluation artifacts can be opened unevenly \cite{liesenfeldOpeningChatGPTTracking2023, whiteModelOpennessFramework2024, bommasani2023}. This literature also grounds critiques of openness as marketing or facade: open releases may expand access without redistributing governance or profits \cite{segerDemocratisingAIMultiple2023}, may claim openness without transparency or public benefit \cite{liesenfeldRethinkingOpenSource2024}, and may introduce safety, misuse, or accountability risks \cite{eirasPositionMidtermRisks2024, segerOpenSourcingHighlyCapable2023, widder2022}. More recent work shifts from model artifacts to the ecosystems around them. Osborne et al. show that Hugging Face downloads are highly concentrated \cite{osborneAICommunityBuilding2024}; Choksi et al. and Castaño et al. find that model artifacts tend to remain relatively static, while durable artifacts and communities move across ephemeral model releases \cite{choksiBriefWondrousLife2025, castanoAnalyzingEvolutionMaintenance2024}; and Longpre et al. identify loosely organized communities that produce derivative models, which are sometimes adopted more widely than their base releases \cite{longpreEconomiesOpenIntelligence2025}. Because AI models are both content and open-ended dual-use tools, Gorwa and Veale argue that model hubs can benefit from external analytic capacity to function as fair and proportionate regulatory access points \cite{gorwaModeratingModelMarketplaces2024}. Together, this work shows that model openness is mediated by the platforms, communities, and artifacts through which models circulate.

Scholarship on participation and political economy of AI extend this infrastructural turn, but remain largely cloud-centered. Widder et al. argue that even maximally open models do not enable meaningful participation when pretraining capacity and inference compute remain concentrated among a small number of firms \cite{widderWhyOpenAI2024, widderOpenBusinessBig2023}; Slater and Masiello, Gansky, and Public AI respond by seeking noncommercial infrastructure for model development and access \cite{masielloWillOpenSource2023, ganskyOpinionArtificialIntelligence2019, publicainetwork2024}. Participatory-AI scholarship clarifies why this matters: broadened access is not the same as redistributed power, especially when publics can inform or use systems without contesting their goals, defaults, or governance, and when openness itself remains a negotiated and value-laden construct that does not guarantee meaningful participation or accountability \cite{birhanePowerPeople2022, corbettPowerPublicParticipation2023, rehakContestingOpennessAI2025, smithReimaginingOpenSource2026}. Suresh et al. suggest that meaningful participation may be more plausible in application-oriented and ``subfloor'' infrastructures than at the general foundation-model layer \cite{suresh2024}, which makes local inference infrastructure especially important. Existing work gestures toward this layer: Nagle et al. show that open models can foster competition among cloud inference providers \cite{nagleLatentRoleOpen2025}, Choksi et al. and Longpre et al. identify local-oriented derivative model communities \cite{choksiBriefWondrousLife2025, longpreEconomiesOpenIntelligence2025}, and Osborne et al. and Linåker et al. show hardware-software optimization as an important axis of open AI collaboration \cite{osborneCharacterisingOpenSource2025, linakerCartographyOpenCollaboration2025}. Yet outside cloud inference, we still lack an account of who maintains local inference infrastructure, who captures value from it, and how its technical dependencies organize participation.

\subsection{Examination of interwoven technical-political development}

Our analysis of \texttt{llama.cpp} proceeds in light of past work examining the interwoven trajectories of technical development and the political systems of power which they interact with. Janet Abbate's history of the Internet treats network architecture as the product of negotiations among military agencies, researchers, standards processes, and users rather than as the inevitable outcome of engineering logic \cite{abbate2000}. Amelia Acker chronicles how Big Tech's platforms serve to commoditize and concentrate control of user data~\cite{acker2025archiving}. In a 2022 FAccT paper, Cooper and Vidan show how \textit{accountability} on the internet was assembled through simultaneous technical and administrative arrangements, moving from mere billing to instrumented policy  \cite{cooperMakingUnaccountableInternet2022}. In the context of AI, Gururaja et al. show how research cultures in AI developed from scientific experimentation to one  shaped by the incentives and infrastructure of Big Tech firms \cite{gururaja2023}. We follow this tradition by treating \texttt{llama.cpp}'s technical record---changed files, backend integrations, model-support paths, conversion scripts, and public documentation of partnerships---as evidence of how local inference is organized and incentivized.

We do this in light of past scholarship on open source software specifically, which shows that firms often participate in open infrastructure not out of altruism, but because it reduces development costs, expands their market, or allows them to commodetize a complementary layer while preserving advantage elsewhere \cite{westHowOpenOpen2003, westChallengesOpenInnovation2006, StrategyLetter2002}. Studies of WebKit and OpenStack show that competitors can cooperate around shared infrastructure when doing so lowers coordination costs or supports adjacent markets in which they continue to compete \cite{teixeiraCollaborationOpensourceArena2014, teixeiraCooperationCompetitorsOpensource2016}. Nguyen-Duc et al. further show that such alliances are often mediated by gatekeepers and selective information sharing around firm-specific advantages \cite{nguyen-ducSoftwareFirmsCollaborate2018}. In open-source AI, Osborne et al. similarly find that corporate participation in shared frameworks often aligns with downstream competitive strategy rather than redistributing governance rights \cite{osborneCharacterisingOpenSource2025, osborneWhyCompaniesDemocratise2024}. This literature gives us a way to read repository activity as strategic participation: contributions are technical changes, but also traces of what actors need the infrastructure to become.

\section{Methods}
\label{sec:methods}
\subsection{Site}
\label{sec:site-background}

\subsubsection{\texttt{llama.cpp} as site of local inference infrastructure}
\label{sec:site-llamacpp}

\texttt{llama.cpp} is an open-source C/C++ inference tool for running large language models on user-owned machines rather than through a cloud API. It loads model weights, manages memory, executes the numerical operations required to generate each token, and exposes command-line and server interfaces for local use. It began in March of 2023 when Georgi Gerganov used \texttt{ggml}---his low-level tensor library for efficient machine-learning inference on the CPU---to run Meta's LLaMA models locally.

We study \texttt{llama.cpp} because it is a widely adopted, hardware-agnostic inference runtime with extensive community uptake and support for many open models, and it underpins numerous downstream tools used for local deployment (e.g., Ollama\footnote{Ollama provides a curated/pullable model registry, making models easier to download and run. Its local inference path depends on \texttt{llama.cpp}. \url{https://github.com/ollama/ollama/}}, LM Studio\footnote{LM Studio is a Desktop GUI for running models locally. In their docs, they write ``LM Studio supports running LLMs on Mac, Windows, and Linux using \texttt{llama.cpp}.'' \url{https://lmstudio.ai/docs/app}}).

\subsubsection{From LLaMA to local inference}
\label{sec:from-llama-to-local-inference}

LLaMA was the first open-weight model series to implement the Chinchilla\footnote{Chinchilla is the name of the model that demonstrated how for a fixed compute budget, model size and training data should be scaled together rather than maximizing parameter count alone} scaling laws \cite{hoffmannTrainingComputeOptimalLarge2022}, resulting in its 13B model outperformed GPT-3 on many benchmarks despite being roughly ten times smaller \cite{touvronLLaMAOpenEfficient2023}. However, Meta's official inference code was written in \texttt{PyTorch} with \texttt{CUDA}, NVIDIA's GPU-acceleration library, presupposing access to relatively expensive NVIDIA hardware\footnote{A sufficiently powerful GPU cost upwards of \$1599 at the time, with current prices closer to \$3399. \url{https://www.tomsguide.com/news/nvidia-geforce-rtx-4090}}.

Gerganov's intervention was to make LLaMA run acceptably on CPUs and system RAM, which most users of computers already own. At inference time, a model repeatedly streams billions of weights from memory to the processor to produce the next token; for many local inference workloads, the time spent moving those weights is more limiting than the processor's raw arithmetic capacity, making CPUs and system RAM a viable alternative to expensive GPUs and VRAM \cite{recasens2025}. \texttt{llama.cpp} reduced this burden by using quantized\footnote{Quantization stores model weights with fewer bits than full-precision formats, reducing memory use and memory-bandwidth demand at some cost to numerical precision and sometimes model quality.} weights, including 4-bit representations, so that a 7-billion-parameter model could fit in ordinary RAM. While the bottleneck of a CPU gets larger as the model size gets larger, CPU inference of smaller models can at times yield even greater performance in local settings than GPU \cite{zhang2025}. As the repository manifesto\footnote{\url{https://github.com/ggml-org/llama.cpp/discussions/205}} states, Gerganov desired to ``bring ideas from the cloud to the device, in the hands of users'' through \texttt{llama.cpp}.

\subsubsection{From independent infrastructure to institutional support}
\label{sec:site-institutional-trajectory}

In June~2023, Gerganov founded \texttt{ggml.ai} as a company to support the library's development\footnote{In the company website, Gerganov states its minimal goals and philosophy, without a clear business model: \url{https://ggml.ai/}}, with backing from Nat Friedman and Daniel Gross through AI Grant\footnote{\url{https://aigrant.com/}}. This gave the project an independent runway, but not a durable institutional home; Section~\ref{section:hf-incentive} returns to the financing structure when explaining why absorption into Hugging Face was an available path.

\begin{figure}[H]
    \centering
    \includegraphics[width=0.8\linewidth]{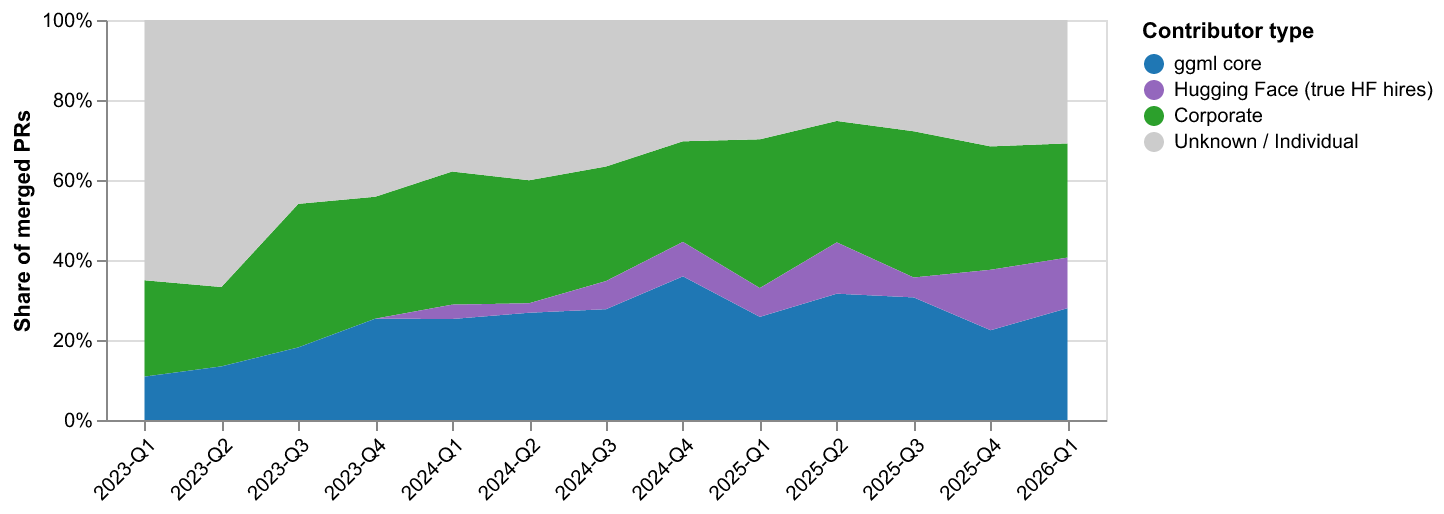}
    \caption{A normalized stacked bar chart showing merged PRs per quarter, where each stack is a contributor type.}
    \Description{Normalized stacked bar chart of merged pull requests per calendar quarter from 2023 through 2026, with each bar divided into segments by contributor type (individual contributors, ggml core team, hardware-making companies, and model-making companies). The individual-contributor segment shrinks from roughly 65 percent of each quarter's bar in the earliest quarters to roughly 30 percent by 2026, while corporate and core-team segments grow correspondingly.}
    \label{fig:share-pr}
\end{figure}

Over time, it drew contributors from model makers, hardware manufacturers, model distributors, and independent developers. Figure \ref{fig:share-pr} shows that individual contributors, who accounted for roughly 65\% of merged pull requests in the repository's earliest quarters, had fallen to approximately 30\% by 2026, their share progressively displaced by corporate contributors and the ggml core team. Corporate affiliations identified in the repository span 47 model-making affiliations and eight hardware-making affiliations.

\subsection{Quantitative analysis}

We collect PRs through GitHub's API and treat merged PRs as evidence of engineering work, while recognizing that they do not directly measure user adoption, revenue, or strategic importance~\cite{kalliamvakouPromisesPerilsMining2014}. For each merged PR, we record the files it changed; GitHub provides at most 300 changed files per PR, so a small number of very large refactors may be incomplete.

We classify each PR by the part of \texttt{llama.cpp} it changes: hardware backends, meaning device-specific code for CPUs, GPUs, and other accelerators; model implementations; conversion tools that translate model releases into files \texttt{llama.cpp} can load; server and example applications; and core inference code. PRs can receive multiple categories. We rely mainly on changed file paths rather than GitHub labels because paths are more stable over time~\cite{herzigItsNotBug2013}, especially after the project separated hardware and model code in June~2024 (PR~\#8006) and later split hardware backends into per-backend directories (PR~\#10256).

For older PRs whose paths predate this clearer structure, we use titles only when paths are insufficient: model names identify model-support work, and CPU instruction names such as AVX, NEON, SVE, and AMX identify CPU-related work when the PR does not also touch GPU backend files. We record whether each classification came from file paths or from a title-based rule.

We also assign PR authors to contributor groups. This is difficult because GitHub profiles are incomplete, people change jobs, and some contributors work for more than one organization~\cite{valievEcosystemlevelDeterminantsSustained2018}. We therefore use several signals in order: membership in the \texttt{ggml-org} GitHub organization; the organization suggested by the author's commit email domain; the author's self-reported \texttt{company} field after normalizing spelling variants and subsidiaries; the raw company text when it cannot be normalized; and \texttt{unknown} when none of these signals is reliable. We manually review conflicts between these signals. The \texttt{ggml-org} check matters after the Hugging Face transition because several long-standing core developers now list Hugging Face as their company; keeping them identifiable as core maintainers prevents them from being confused with other Hugging Face contributors. These assignments are thus conservative, especially for former employees, and dual-affiliated contributors.

\subsection{Qualitative analysis}

Quantitative PR patterns identify where activity concentrates, but they do not by themselves show why a firm or contributor acts. For each significant corporate actor, we therefore assemble a dossier of public texts and read it alongside the PR data: blog posts, product releases, partnership listings, vendor documentation, consequential PRs, contribution guidelines, acquisition announcements, contributor reflections, and selected issue threads. We source these materials from the repository itself, company websites, personal technical blogs, public company pages, and media coverage, prioritizing materials that explicitly reference \texttt{llama.cpp}, \texttt{ggml}, or the relevant backend, model, or integration.

We use these materials to interpret corporate strategy, identify organizational relationships not obvious from PR counts alone, and read infrastructure integrations as qualitative evidence. In the Hugging Face analysis, for example, PRs adding \texttt{convert\_hf\_to\_gguf.py}, Hugging Face Hub loading, and features that download or load Hub models directly help show how distribution, conversion, and execution become linked in practice; acquisition announcements, contributor blogs, and \texttt{llama.cpp}'s contribution guidelines help interpret the governance significance of those links.

\section{Findings}

\subsection{The cost of hardware expansion beyond the CPU}
\subsubsection{From CPU-first to a heterogeneous hardware platform}
\label{sec:heterogeneous-hardware-platform}

\begin{figure}[H]
    \centering
    \includegraphics[width=0.8\linewidth]{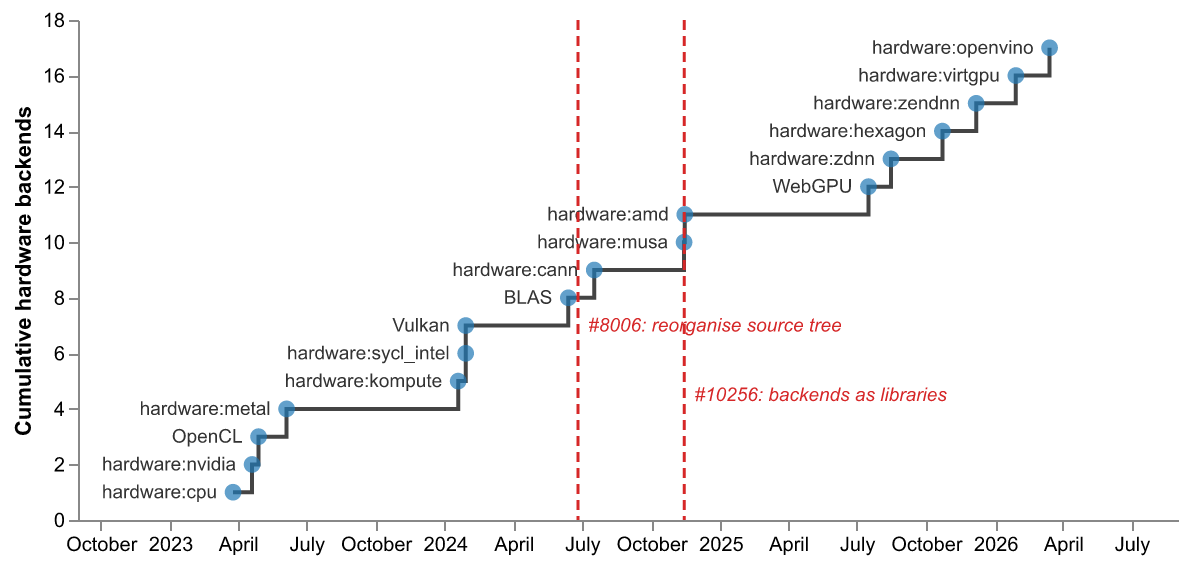}
    \caption{Backends supported on \texttt{llama.cpp} over time}
    \Description{Timeline chart showing the cumulative number of distinct hardware backends supported by llama.cpp from March 2023 through March 2026, rising from a single CPU-only backend at founding to 17 distinct backends by 2026, with new backends appearing at an increasing rate after mid-2024.}
    \label{fig:backend-timeline}
\end{figure}

From its CPU-centered origins, \texttt{llama.cpp} has expanded to support 17 distinct hardware backends as of March 2026, a trajectory shaped by two structural interventions that lowered the barrier to entry of participation. PR \#8006\footnote{\url{https://github.com/ggml-org/llama.cpp/pull/8006}} decoupled hardware and model code into separate directories, clarifying where backend contributions belong; PR \#10256 introduced a more extensive refactoring of the backend architecture\footnote{\url{https://github.com/ggml-org/llama.cpp/pull/10256}}, reorganizing backend implementations into discrete components with clearer separation in the build system. The stated objective was to support a unified \texttt{llama.cpp} distribution capable of targeting heterogeneous hardware through configurable backend inclusion, rather than requiring entirely separate builds per platform. An important consequence of this change is that backend development became more decoupled: contributors were now able to work on, compile, and test individual backends with reduced dependence on the full \texttt{ggml} compilation pipeline. Collectively, these changes invited hardware other than the CPU to become more easily integrated into \texttt{llama.cpp}, as evidenced by the growth shown in figure \ref{fig:backend-timeline}.

\begin{figure}[H]
    \centering
    \includegraphics[width=0.8\linewidth]{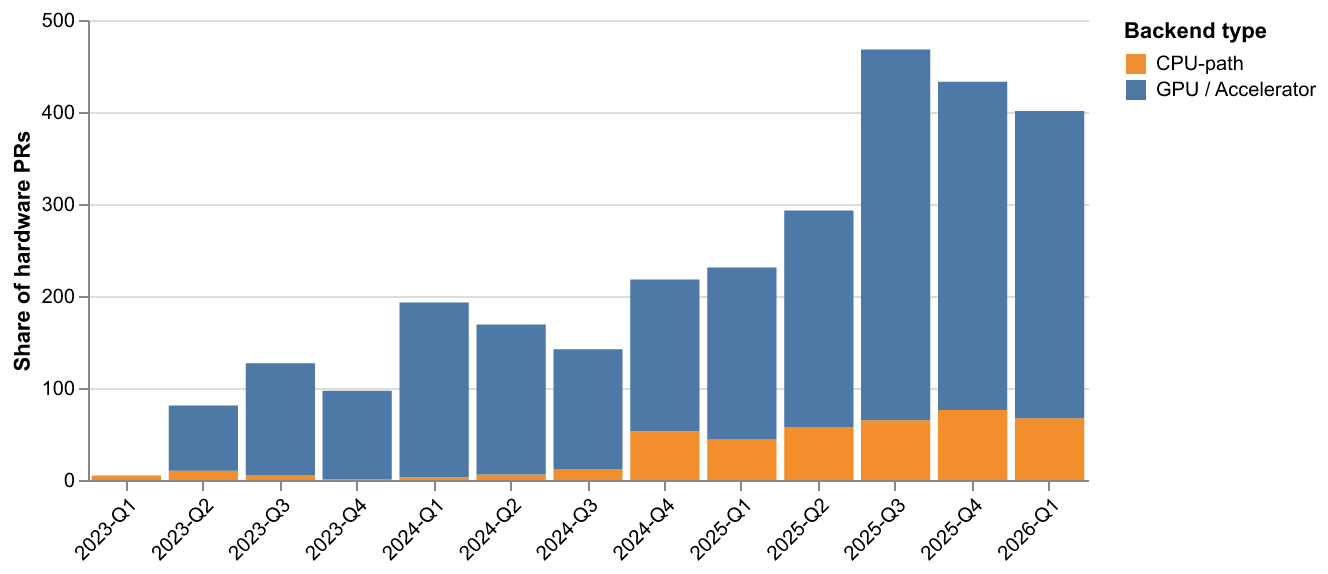}
    \caption{A stacked bar chart of the hardware-related PRs, where one stack represents CPU-related PRs, and the other stack represents accelerator-related PRs. Despite being CPU only in 2023-Q1, accelerator-related PRs dominate for the rest of the repository's history.}
    \Description{Stacked bar chart of hardware-related merged pull requests per quarter, split into a CPU-related segment and an accelerator/GPU-related segment. Only 2023-Q1 is entirely CPU; from 2023-Q2 onward the accelerator segment is the majority of hardware PRs in every quarter through 2026, and the CPU segment never regains a majority share.}
    \label{fig:cpu-vs-gpu}
\end{figure}

Despite this expansion making \texttt{llama.cpp} more accessible to a wider range of hardware contributors, GPU and accelerator-related pull requests have constituted the overwhelming majority of hardware contributions since the project's earliest quarters---a pattern that sits in tension with the project's own contribution guidelines. The documentation instructs contributors to ``focus on CPU support only in the initial PR unless you have a good reason not to,'' a norm that frames CPU compatibility as the baseline from which GPU support may optionally extend. Yet as Figure \ref{fig:cpu-vs-gpu} shows, GPU and accelerator PRs dominated hardware contributions from the outset, with CPU-path PRs never recovering a majority share in any quarter. One interpretation is that the CPU-first guideline has functioned less as a description of practice and more as a floor: a minimum accessibility requirement that GPU-focused contributors satisfy before moving on to their primary target. Another is that the guideline reflects the project's original identity more than its current priorities, and that the contributor base has reoriented the project's hardware focus from the bottom up. The data alone cannot adjudicate between these readings, but either way, the pattern suggests a divergence from the project's original CPU-first emphasis in practice.

\subsubsection{Hardware support as expansion and enclosure: vendor strategies in \texttt{llama.cpp}}
\label{sec:vendor-strategies}

\begin{figure}[H]
    \centering
    \includegraphics[width=0.8\linewidth]{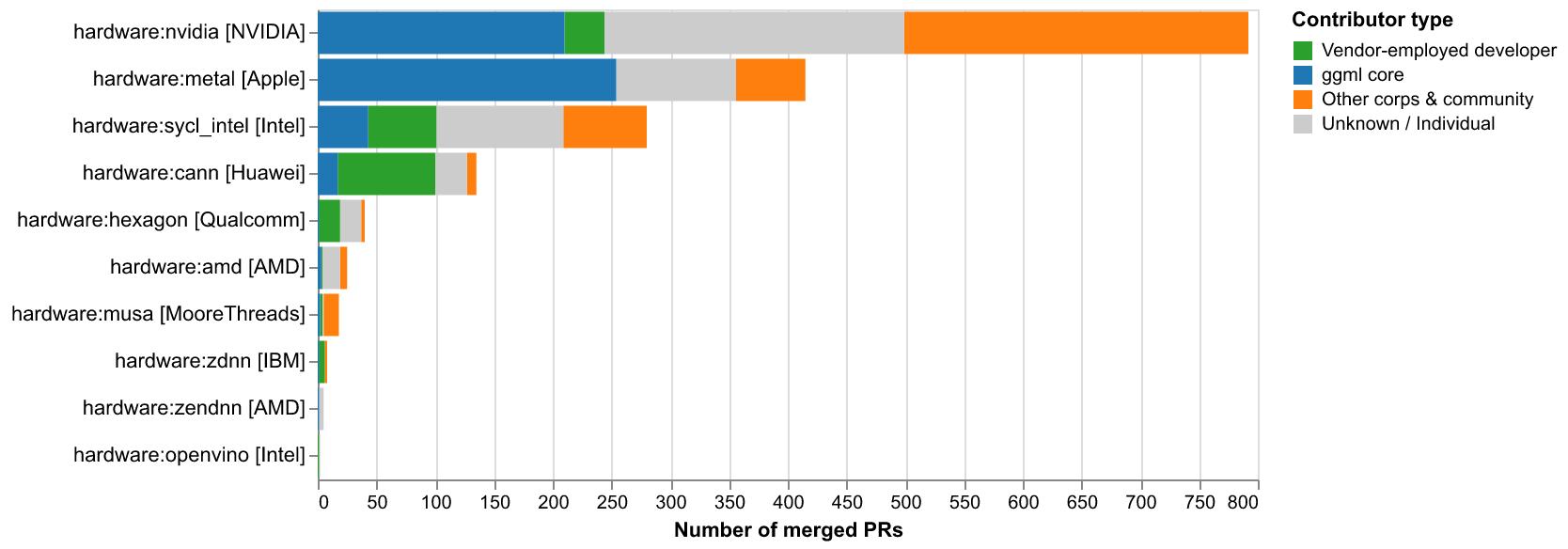}
    \caption{A stacked bar chart of the backend-related PRs, where each stack is the contributor type. Only backends with a specific vendor are listed here.}
    \Description{Stacked bar chart of merged pull requests per vendor-specific backend (NVIDIA CUDA, Apple Metal, Intel SYCL, Huawei CANN, Qualcomm Hexagon, AMD), with each bar's segments colored by contributor type. CUDA has the largest total, at roughly 800 PRs, nearly double Apple Metal's roughly 420; Intel, Huawei, Qualcomm, and AMD follow at progressively smaller totals.}
    \label{fig:nvidia-dominance}
\end{figure}

The distribution of vendor-specific backend contributions reflects the distinctive geography of the local inference market. As Figure \ref{fig:nvidia-dominance} shows, the NVIDIA CUDA backend commands the highest PR volume by a substantial margin, at roughly 800 merged PRs---nearly double the second-ranked Apple Metal backend at approximately 420. Below Apple, Intel's SYCL backend (\textasciitilde280 PRs) and Huawei's CANN (\textasciitilde130) register meaningful levels of investment, followed by Qualcomm's Hexagon and AMD at smaller but non-negligible counts. A demand-side proxy from GitHub release downloads broadly mirrors this hierarchy while also tempering it\footnote{Data sourced from \url{https://tooomm.github.io/github-release-stats/?username=ggml-org&repository=llama.cpp}}: among backend-labeled release assets, CUDA accounts for 24.3\% of downloads, followed by Apple/Metal at 8.7\%, Huawei CANN at 6.3\%, AMD at 5.5\%, Intel at 4.5\%, and Qualcomm/Adreno at 0.5\%. While PR volume measures contributor effort rather than end-user adoption, and release downloads are themselves an imperfect proxy, the convergence between the two measures is nonetheless notable: AMD, Apple, Huawei, Qualcomm, and Intel are either marginal or structurally absent from the cloud GPU market\footnote{According to Silicon Analysts, NVIDIA holds 80-90\% of AI accelerator revenue share, with AMD at roughly 7\% and Intel at 1.5\%. \url{https://siliconanalysts.com/analysis/nvidia-ai-accelerator-market-share-2024-2026}}, yet each is represented in both upstream backend work and downstream binary demand for \texttt{llama.cpp}. This pattern suggests that local inference infrastructure has become a competitive surface where hardware vendors who cannot contest NVIDIA's dominance in the cloud have a meaningful stake, and where the hierarchy of participation looks materially different from the one that governs data center AI.

\begin{figure}[H]
    \centering
    \includegraphics[width=0.8\linewidth]{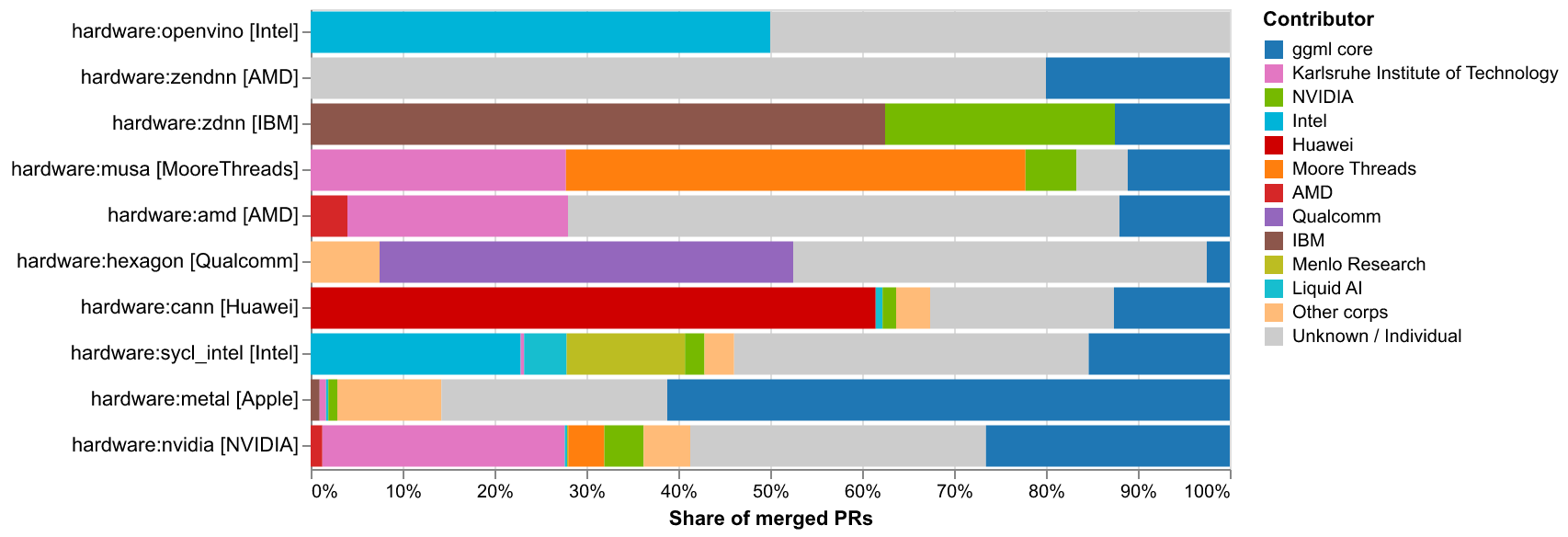}
    \caption{A normalized stacked bar chart of the backend-related PRs, where each stack is the contributor type. Only backends with a specific vendor are listed here. Corporations and contributors are further divided up into specific, named stacks.}
    \Description{Normalized stacked bar chart of the CUDA backend's merged pull requests, broken into named contributor segments rather than broad categories. The ggml core team and other independent/corporate contributors make up most of the bar; NVIDIA's own engineers (shown in green) are a minority segment, smaller than the top individual contributor (a Karlsruhe Institute of Technology researcher) and comparable to or smaller than AMD's and Moore Threads' segments.}
    \label{fig:nvidia-dominance-norm}
\end{figure}

Despite NVIDIA's overwhelming share of backend PR volume, the CUDA codebase is maintained primarily by the \texttt{ggml} core team, and other corporate or independent contributors rather than NVIDIA itself---a gap that AMD and Moore Threads\footnote{Moore Threads may be a leseser known corporation to some, but it is a Chinese company specializing in making graphics cards, founded by former vice president of NVIDIA's Chinese branch. \url{https://www.scmp.com/business/banking-finance/article/3335269/moore-threads-stock-surge-fivefold-shanghai-debut-investors-rush-ai-chipmaker}} have exploited through interoperability. As Figure \ref{fig:nvidia-dominance-norm} shows, NVIDIA's own engineers (in green) are a minority presence; the dominant individual contributor is a researcher from the Karlsruhe Institute of Technology operating in an individual capacity, with AMD and Moore Threads following behind. Rather than building independent backends, both AMD and Moore Threads have developed CUDA-compatible programming languages---ROCm (AMD) and MUSA (Moore Threads)---designed to map directly onto CUDA's API. Their contributions to llama.cpp's CUDA codebase consist largely of thin compatibility layers, such as header files that substitute CUDA function calls with their ROCm or MUSA equivalents at compile time. This approach requires minimal original engineering effort while securing compatibility with the most actively maintained backend in the repository. This dynamic illustrates a broader tension in local inference infrastructure: while CUDA itself is proprietary, the openness of the \texttt{llama.cpp} codebase enables broad participation and creates opportunities for strategic appropriation.

\begin{figure}[H]
    \centering
    \includegraphics[width=0.8\linewidth]{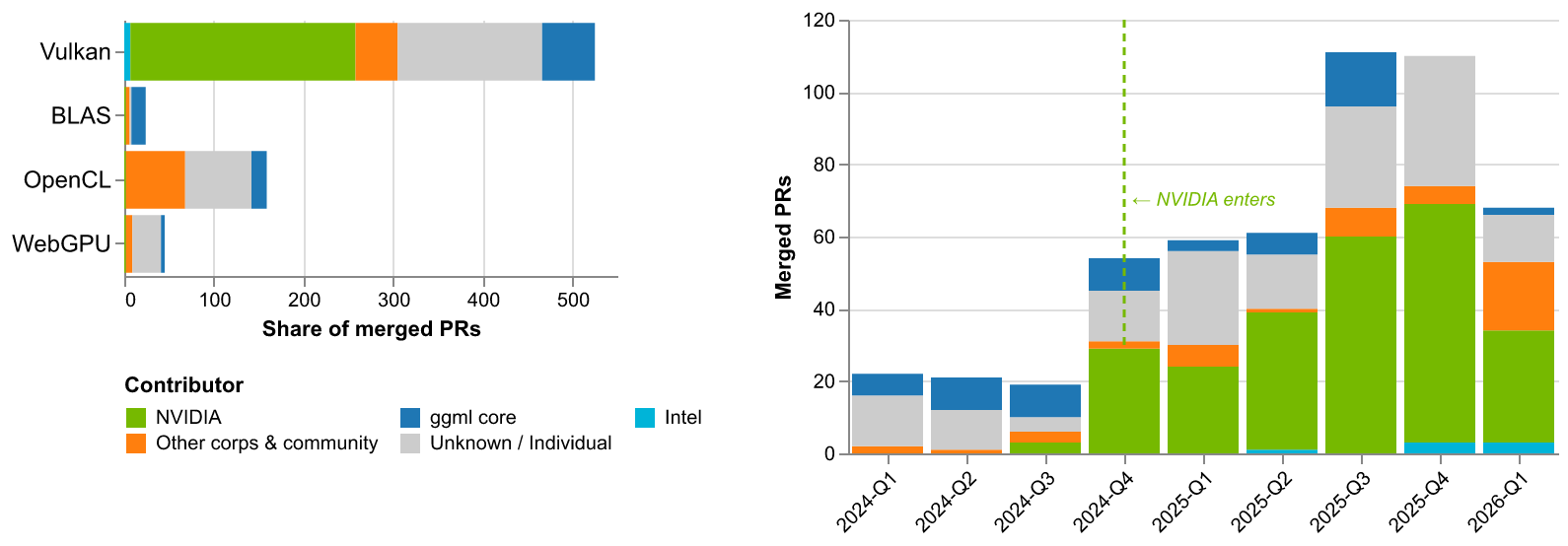}
    \caption{On the left, a stacked bar chart of PRs for cross-platform backends. On the right, a stacked bar chart of PRs for the Vulkan cross-platform backend by quarter.}
    \Description{Two-panel chart. The left panel is a stacked bar chart comparing total merged pull requests across the three open, vendor-neutral cross-platform backends (Vulkan, OpenCL, WebGPU), with Vulkan receiving substantially more activity and corporate investment than the other two. The right panel is a stacked bar chart of Vulkan pull requests by quarter, colored by contributor type, showing NVIDIA's share of Vulkan activity surging from Q4 2024 onward to roughly 50 percent of all Vulkan PRs by 2026.}
    \label{fig:nvidia-capture}
\end{figure}

CUDA's dominance is representative of NVIDIA's hardware market share rather than sustained upstream engineering investment in the backend; nevertheless, NVIDIA is extending its influence into adjacent layers by actively shaping the Vulkan backend---nominally the open, vendor-neutral cross-platform backend for GPU acceleration---through contributions optimized for its own hardware. As Figure \ref{fig:nvidia-capture} shows, NVIDIA's Vulkan PR volume surged from Q4 2024 onward, reaching approximately 50\% of all Vulkan activity by 2026; OpenCL and WebGPU, the other open-standard backends, received no meaningful corporate investment. The character of NVIDIA's Vulkan contributions is instructive: PR \#10206\footnote{https://github.com/ggml-org/llama.cpp/pull/10206}, authored by an NVIDIA engineer, adds support for \texttt{VK\_NV\_cooperative\_matrix2}---a vendor-specific Vulkan extension, denoted by the NV prefix, that is exclusive to NVIDIA drivers and hardware. Another prominent contributor's immediate response was to ask whether a cross-vendor \texttt{VK\_KHR} implementation could follow, implicitly acknowledging that the contribution tilts a shared backend toward a proprietary feature set. Taken together with the AMD and Moore Threads dynamic described above, the Vulkan case illustrates the inverse strategy: where AMD and Moore Threads piggyback on CUDA's dominance through interoperability, NVIDIA works to ensure that even the open alternative converges on its hardware. Local inference infrastructure, in this sense, reflects a tension between two overlapping strategies: challengers gaining footholds through interoperability, and incumbents consolidating existing advantages.

\subsection{The cost of model expansion beyond LLaMA}
\subsubsection{From LLaMA to a heterogeneous model platform}
\label{sec:hetero-model}

\texttt{llama.cpp}'s model support has expanded steadily since its founding, growing from a single architecture to 104 distinct model families across 2.5 years. Here, a model family refers to a related set of model releases that require identifiable support in \texttt{llama.cpp}, while a model architecture refers to the underlying design that determines how the runtime must load weights and execute the model. With 36 canonical model architectures present in \texttt{llama.cpp}, Chinese companies represent the single largest category, nearing double American Big Tech companies (18) and academic institutions (20). The Chinese wave is also recent and intensifying: 9 new Chinese models in 2025-Q4 alone, and 7 more in 2026-Q1. This reflects the rapid proliferation of Chinese open-weight model labs all releasing distinct models in close succession, which has also been documented in Longpre et al.'s findings on Hugging Face \cite{longpreEconomiesOpenIntelligence2025}.  Beyond the US--China axis, a set of smaller national presences---UAE, South Korean, Japanese, Canadian, Israeli, and UK companies---each account for one to two models, mostly post-2024, suggesting that geographic diversification is a recent and ongoing phenomenon.

\begin{figure}[H]
    \centering
    \includegraphics[width=0.8\linewidth]{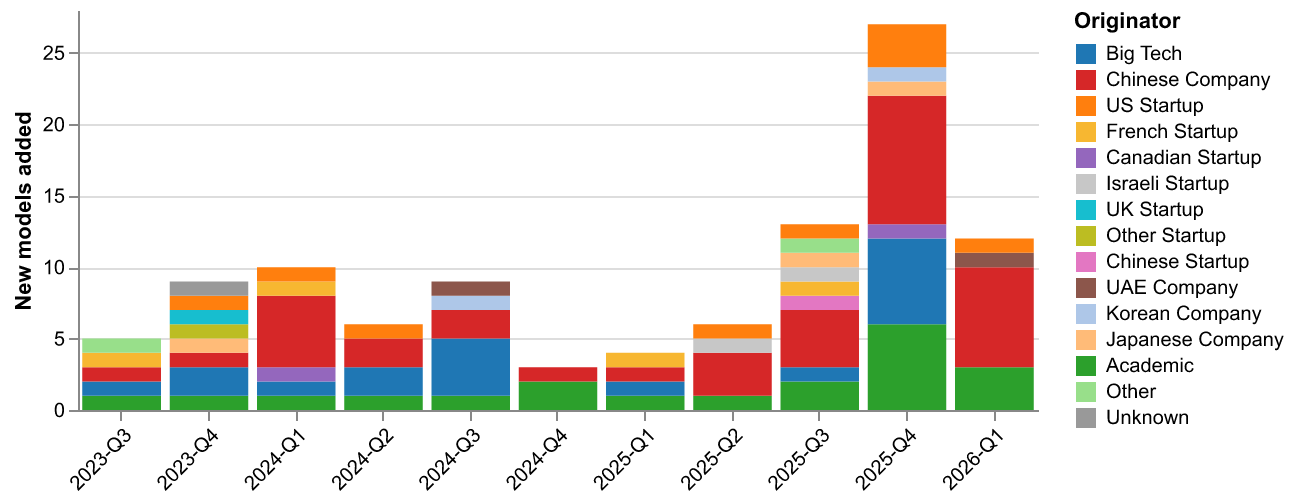}
    \caption{New model architectures added over time}
    \Description{Timeline chart showing the cumulative number of distinct model architectures added to llama.cpp from March 2023 through March 2026, reaching 36 canonical architectures by the end of the period, with new-architecture additions from Chinese companies clustering heavily in late 2025 and early 2026 (9 in 2025-Q4 and 7 in 2026-Q1).}
    \label{fig:model-ecosystem}
\end{figure}

Interestingly, unlike hardware vendors who participate extensively in shaping backends, model companies rarely maintain the code needed to load and run their own models in \texttt{llama.cpp}. Across the 69 model families with five or more merged PRs, model owners contribute a median of zero PRs to their own implementations, accounting in aggregate for just 6.7\% of model-related activity. The asymmetry in maintenance effort is further reflected in PR volume: hardware backends attract a median of 40 merged PRs each (max 792), while model families receive a median of 5 (max 32), suggesting that model support is characterized by sparse, one-off contributions rather than sustained maintenance. The dominant maintainers of model support are instead \texttt{ggml} core contributors (34.4\%) and community members of unknown affiliation (30.2\%), with only seven companies---Hugging Face, IBM, Microsoft, NVIDIA, LiquidAI, Tencent, and Google---contributing any PRs at all. The five companies sustaining more than two PRs share a common characteristic: each has a structural relationship with \texttt{ggml} or \texttt{llama.cpp} that goes beyond general interest in local inference. NVIDIA is listed as an official partner on the \texttt{ggml} GitHub organization page\footnote{NVIDIA uses the \texttt{llama.cpp} backend to boast its superior performance over Apple chips in local model inference: \url{https://www.pcworld.com/article/2916928/nvidia-rtx-5090-outperforms-amd-and-apple-running-local-openai-language-models.html}. See also \url{https://github.com/ggml-org}.}; Liquid AI produces edge-optimized ultra-small models and explicitly names \texttt{llama.cpp} as a recommended runtime in its release materials\footnote{\url{https://www.liquid.ai/blog/liquid-foundation-models-v2-our-second-series-of-generative-ai-models}.}; IBM has a contributor embedded within ggml-org specifically to support Granite and IBM's mainframe backend\footnote{Aaron Teo is both an IBM employee and a member of ggml-org on GitHub. While it is unclear how this arrangement came to be, Teo states in their blog that he decided to work on \texttt{llama.cpp} because of its first-class support for CPUs: \url{https://medium.com/@taronaeo/a-year-ago-and-now-progress-of-llama-cpp-on-mainframes-s390x-e99061abc43b}.}; and Hugging Face, which now owns \texttt{ggml} outright, maintains a consistent presence across nearly all model families\footnote{Gerganov, in the official announcement of the HF-ggml merge, mentions two HF engineers \texttt{ngxson} and \texttt{allozaur} who "contributed several core functionalities to \texttt{ggml} and \texttt{llama.cpp}, among other contributions.'' \url{https://github.com/ggml-org/llama.cpp/discussions/19759}}. Model owner participation, where it exists, is concentrated among companies with formal or strategic ties to \texttt{ggml}. This pattern suggests that the local model ecosystem is decoupled from its upstream producers: models are widely distributed, but sustained integration work in \texttt{llama.cpp} concentrates in a small set of actors with strategic stakes in the local inference layer, rather than the model layer itself.

\subsubsection{Understanding Hugging Face's Incentives} \label{section:hf-incentive}

In February~2026, \texttt{ggml.ai} joined Hugging Face, with Gerganov and the core team becoming full-time Hugging Face employees\footnote{Announcement of the merge on GitHub, mainly authored by Gerganov. \url{https://github.com/ggml-org/llama.cpp/discussions/19759}}. The announcement presented the move as a way to secure long-term support for local AI while keeping \texttt{ggml} and \texttt{llama.cpp} open-source under their existing MIT licenses. It also described the transition as a formalization of work already underway between the two organizations, including GGUF format work, conversion from Hugging Face tensor and metadata conventions into GGUF\footnote{GGUF is a format for storing model weights together with metadata needed to make sense of the weights. See \url{https://huggingface.co/docs/hub/gguf} for more details.}, \texttt{transformers}-based metadata handling, built-in Hub loading, and deeper integration with the Hugging Face Hub\footnote{Announcement of the merge on Hugging Face, authored mostly by Hugging Face and some from ggml-org. \url{https://huggingface.co/blog/ggml-joins-hf}}. Thus, by the end of the period studied here, \texttt{llama.cpp} remained an open repository, but its core maintenance work had moved from an independent company into a larger model-distribution platform.

Hugging Face's position is best understood as that of an infrastructural intermediary. Founded in 2016 as a consumer chatbot, the company pivoted after open-sourcing \texttt{transformers}, whose adoption exceeded that of the original product\footnote{\url{https://research.contrary.com/company/hugging-face}}. Its subsequent business model paired open libraries and public model hosting with enterprise contracts and managed hosting; by 2023, Hugging Face reportedly generated approximately \$70M in annual recurring revenue, largely from closed managed versions of its platform\footnote{\url{https://sacra.com/c/hugging-face/}}. The relevant incentive is therefore not to close the open-model ecosystem, but to remain the default site through which open models are discovered, documented, converted, and deployed.

The company's financing underscores the value of this intermediary position. Hugging Face's \$235 million Series D included Google, Amazon, Nvidia, Intel, AMD, Qualcomm, IBM, Salesforce, and Sound Ventures\footnote{\url{https://techcrunch.com/2023/08/24/hugging-face-raises-235m-from-investors-including-salesforce-and-nvidia/}}: firms that compete across cloud, hardware, enterprise software, and model markets, but share an interest in a common distribution layer that no one of them fully owns. Its reported rejection of a \$500 million Nvidia investment offer is consistent with this position---declining the deal to avoid a single dominant investor able to influence its decisions \cite{heikkila2026huggingface}. Hugging Face's power does not depend on subordination to one sponsor; it depends on being sufficiently neutral that competing actors can treat the platform as shared infrastructure.

Neutrality, however, does not preclude concentration. Hugging Face's acquisitions have tended to incorporate interfaces that make open models usable: Gradio for demonstrations\footnote{\url{https://huggingface.co/blog/gradio-joins-hf}}, Argilla for data labeling and curation\footnote{\url{https://argilla.io/blog/argilla-joins-hugggingface/}}, Pollen Robotics for embodied data and hardware\footnote{\url{https://techcrunch.com/2025/04/14/hugging-face-buys-a-humanoid-robotics-startup/}, \url{https://huggingface.co/blog/hugging-face-pollen-robotics-acquisition}}, and now \texttt{ggml.ai} for local execution. These tools often remain open and broadly useful, which is precisely why they can become infrastructurally consequential. The issue is not that Hugging Face prevents openness, but that it increasingly organizes the practical conditions under which openness is operationalized: where models are found, how they are documented, how they are converted, and how they are run.

The funding trajectory of \texttt{ggml.ai} further clarifies this dynamic. The backing from Daniel Gross and Nat Friedman (explored in section \ref{sec:site-institutional-trajectory}) supplied an independent runway for \texttt{llama.cpp}, but it did not create a durable institution capable of sustaining long-term maintenance, roadmap coordination, and full-time stewardship. The terms of investment are significant because a no-cap, no-discount MFN SAFE—a funding strategy developed by YCombinator\footnote{\url{https://www.ycombinator.com/documents}} for uncertain-stage fundraising—gives early backers neither a valuation ceiling nor a discounted conversion price; the MFN clause protects only against later investors receiving more favorable terms. In an acqui-hire-like transition, which is characteristic of Hugging Face's acquisition of \texttt{ggml.ai}, unconverted SAFE holders can be weakly positioned because much of the transaction value may be organized as employment, retention, or team-integration value rather than as acquisition consideration paid through a mature capital stack. The significance of this financing is not that Hugging Face's acquisition was predetermined, but that formally open infrastructure still required institutional support, and the available support mechanisms made absorption into a larger platform comparatively easy.

\texttt{llama.cpp} decentralizes where inference happens, but not necessarily how runnable models are found, prepared, and prioritized. It moves computation outward onto user-owned devices, reducing dependence on cloud inference providers. Yet this shift leaves intact the steps that determine which models appear usable in the first place: where users discover model repositories, who coordinates support for new releases, which metadata conventions describe the weights, which scripts convert those weights into local files, which popularity signals guide attention, and which built-in loading options point users toward a hub. In this configuration, Hugging Face does not need to prevent local inference to exercise infrastructural power; it can support local inference while organizing the pathways through which local inference becomes practical. The limit of openness, in this case, is that access remains mediated by centralized infrastructures for making models findable, compatible, and usable.

Table~\ref{tab:local-inference-possibilities-pitfalls} summarizes the paper's findings on the possibilities and pitfalls of local AI inference infrastructure, using \texttt{llama.cpp} as a case study.

\begin{table*}[t]
\centering
\caption{Possibilities and pitfalls of local AI inference infrastructure}
\label{tab:local-inference-possibilities-pitfalls}
\small
\begin{tabularx}{\textwidth}{>{\raggedright\arraybackslash}p{3cm} >{\raggedright\arraybackslash}X >{\raggedright\arraybackslash}X}
\toprule
\textbf{Layer} & \textbf{Possibility} & \textbf{Pitfall} \\
\midrule
Execution and access & Local devices let users run models outside provider-owned APIs, reducing dependence on metered cloud access. (See section~\ref{sec:from-llama-to-local-inference}) & Practical use still depends on upstream model providers, and maintenance work to make consumer hardware compatible. (See section~\ref{sec:open-recapture}) \\
\addlinespace
Hardware and device support & Device-specific code makes local AI usable across CPUs, GPUs, mobile chips, and vendor-specific accelerators. (See section~\ref{sec:heterogeneous-hardware-platform}) & Open device-support work can become a site where incumbents steer cross-platform backends toward their own hardware. (See section~\ref{sec:vendor-strategies}) \\
\addlinespace
Model support & Broad model support makes many open-weight model families runnable through a shared local runtime. (See section~\ref{sec:hetero-model}) & Model owners rarely maintain their own local implementations, leaving integration work to core maintainers, community contributors, and a small set of strategically positioned firms. (See section~\ref{sec:hetero-model}) \\
\addlinespace
Distribution and conversion & Model hubs, common formats, and conversion scripts lower friction for finding, preparing, and running models locally. (See section~\ref{section:hf-incentive}) & Convenience can harden into dependency when one platform's conventions become the default path for distribution, conversion, and runtime loading. (See section~\ref{section:hf-incentive}) \\
\bottomrule
\end{tabularx}
\end{table*}

\section{Discussion}

\subsection{The promise and peril of local AI: broader participation and yet new forms of capture} \label{sec:open-recapture}

Debates over participation in AI increasingly turn on what participation is supposed to mean: access to model artifacts, the ability to use and modify systems, or power to contest the goals, defaults, and institutions through which AI is governed \cite{birhanePowerPeople2022, corbettPowerPublicParticipation2023, rehakContestingOpennessAI2025}. This paper intervenes in that debate by showing that local inference expands participation at the point of execution while relocating capture into the infrastructures that make execution possible. \texttt{llama.cpp} substantially lowers the practical threshold for participation: users can run models on their own machines, independent developers can add model or device support, and hardware vendors outside the dominant cloud GPU market can make their devices usable for contemporary AI. This is precisely the kind of application-oriented or ``subfloor'' site where Suresh et al. suggest meaningful participation may be more plausible than at the foundation-model layer \cite{suresh2024}. It also matches how downstream users understand openness---through reliability, privacy, local control, experimentation, and the ability to assemble usable systems from community-maintained tools \cite{leeOpenAIWild2026}. Yet participation remains unevenly distributed across layers. Individual contributors and users gain access to execution and adaptation, while core maintainers, hardware vendors, and model distributors retain disproportionate influence over download defaults, device compatibility, the timing of support for new model releases, and the ongoing work of reviewing PRs and fixing breakage. Local inference therefore raises the floor of participation without necessarily raising its ceiling.

Corporate participation in \texttt{llama.cpp} is best understood as strategic positioning in the hardware markets opened by local inference, not simply as generic support for open AI. Building on Osborne et al.'s account of open source co-opetition, which shows that firms collaborate in shared AI frameworks while preserving downstream competitive advantage \cite{osborneCharacterisingOpenSource2025}, our analysis shows how this logic travels into consumer-oriented inference infrastructure. A cross-platform inference tool lowers the cost of making many devices AI-capable, while firms use device-support contributions to make their own hardware legible, performant, and compatible. This also specifies Linåker et al.'s finding that open LLM collaboration extends beyond models to frameworks, datasets, and tooling: in \texttt{llama.cpp}, the strategically important artifact is the inference layer where models meet hardware \cite{linakerCartographyOpenCollaboration2025}. Strategic positioning takes two forms. Challenger firms such as AMD and Moore Threads pursue CUDA compatibility, aligning their hardware with the dominant software stack rather than displacing it. Incumbents such as NVIDIA, by contrast, shape nominally cross-platform backends such as Vulkan through vendor-specific optimizations that preserve their performance advantages. Interoperability, in this setting, means concrete compatibility work with competitive consequences: challengers piggyback on dominant standards, while incumbents bend open paths back toward their own hardware.

The competitive gains of open models are real, but our central finding is that they are mediated by the infrastructures through which open models become runnable. Consistent with Nagle et al., openness can lower the marginal cost of inference and intensify competition \cite{nagleLatentRoleOpen2025}; \texttt{llama.cpp} shows that these gains pass through CUDA and similar accelerator software, scripts that convert hub-hosted weights into files \texttt{llama.cpp} can load, and model-distribution platforms such as Hugging Face. This qualification extends political economy critiques of openness and concentration. Widder et al. argue that openness does not ensure meaningful redistribution of power when the resources required to build and deploy AI systems remain concentrated \cite{widderOpenBusinessBig2023}, while Narechania et al. locate this concentration across interconnected layers of the AI stack \cite{narechaniaAntimonopolyApproachGoverning2024}. In the local domain, inference can reduce dependence on hyperscaler infrastructure while introducing new centralization around who coordinates new model releases and where users fetch and load those models. Hugging Face's integration across model hosting, model-file conversion, and loading features built into local runtimes positions it as an infrastructural intermediary approaching an ``obligatory passage point'' \cite{callonElementsSociologyTranslation1984}. These mechanisms do not nullify broader participation, but they condition it: users and contributors gain new paths to run and adapt models, while control over findability, compatibility, release timing, and defaults accumulates in fewer infrastructural hands.

\subsection{Policy implications: preserving openness and localness beyond the model itself}
\label{sec:policy}

Federal documents on open models define openness as a property of the model itself. NTIA's 2024 Request for Comment framed it as a definitional and risk-governance problem \cite{ntiaDualUseFoundation2024}; Meta emphasized innovation, competition, safety research, and U.S. leadership \cite{metaNTIARFC2024}; the Open Source Initiative identified broad support for open models across the docket \cite{osiCompellingResponses2024}; and the Johns Hopkins Center for Health Security offered a narrower biosecurity caution \cite{johnsHopkinsNTIARFC2024}. NTIA's final report synthesized this as cautious openness, supporting open models while building evidence for possible future intervention \cite{ntiaSupportsOpenModels2024}. Mozilla's submission was the only one to argue that meaningful openness extends beyond weights to documentation, data, code, licensing, governance, and infrastructure \cite{mozillaNTIARFC2024}. \texttt{llama.cpp} bears that out. Model weights are useful only through model registries, metadata that tells runtimes how to interpret them, loadable weight formats, conversion scripts, and inference runtimes. The policy question is not whether weights are available but whether the infrastructure around them stays portable, interoperable, and independently governable.

The evidence from \texttt{llama.cpp} clarifies three concrete problems and the mechanisms by which they could be addressed. The first is model-owner responsibility. Model owners contribute a median of zero pull requests to their own implementations in the repository, while hardware vendors show a median of forty PRs per backend. The labor of making models runnable falls on core maintainers, independent contributors, and a small set of strategically positioned firms. The second is format dependency and default loading paths. Hugging Face's pull requests added the Hugging Face Hub loading feature, \texttt{convert\_hf\_to\_gguf.py}, and metadata integration that links distribution, conversion, and execution. Users who run models through these pathways depend on Hugging Face conventions: the model registries they discover, the conversion scripts they run, and the default loading options that point toward one hub over another. The third is vendor-specific capture of cross-platform backends. NVIDIA contributes vendor-specific Vulkan extensions while AMD and Moore Threads contribute ROCm and MUSA header files that map CUDA function calls to their own APIs. Our empirical evidence on NVIDIA shifting from near-zero to roughly 50\% of Vulkan PRs by 2026 shows how an incumbent can steer a shared backend toward its own hardware.

These three problems map to three existing policy mechanisms. The FTC's 6(b) investigative authority documented cloud compute dependencies in a pair of reports on AI partnerships and cloud computing infrastructure \cite{ftcBehind6B2025, ftcCloudRFI2023}. The same authority could map format dependencies in local inference. A 6(b) report on model-distribution infrastructure would document format adoption patterns, how Hugging Face loading flows into runtime defaults, and which backends carry vendor-specific extensions. The data would allow the FTC to evaluate whether open-model infrastructure remains contestable when the code is publicly available. This is not a hypothetical exercise. The repository shows three concrete dependency chains: format lock-in, in which one hub's metadata conventions become the de facto standard through conversion scripts and loading options that users encounter before alternatives; backend capture, in which vendor-specific extensions alter the competitive trajectory of cross-platform backends; and the asymmetric maintenance burden, in which model owners who release weights rely on a small set of unrelated firms to maintain the code that runs them.

The second mechanism is model certification. The FTC has used rulemaking to require that products demonstrate compatibility or interoperability across competitors. A comparable rule for open models would require that model releases exceeding a parameter threshold include tested inference implementations for at least one non-CUDA backend, alongside conversion scripts for standard formats. This shifts the labor of making models runnable from community infrastructure to the entities releasing the models. The policy does not require model owners to build proprietary inference tools. It requires them to contribute to the shared infrastructure that the research shows they currently do not maintain. IBM's embedded contributor on Granite and LiquidAI's explicit recommendation of \texttt{llama.cpp} in their release materials demonstrate that model owners who care about local inference can contribute to cross-vendor runtimes without building parallel stacks.

The third mechanism is public funding for inference maintenance. AI Now and Data \& Society's critique of NAIRR warns that public infrastructure depends on the incumbent firms it aims to counter \cite{aiNowDataSocietyDemocratizeAI2021}. The \texttt{ggml.ai} acquisition illustrates why this problem applies to inference infrastructure as well. Its early backing came through a no-cap, no-discount MFN SAFE. The structure provided runway. It did not fund full-time maintenance, roadmap coordination, or release review indefinitely. When those needs became unavoidable, the path of least resistance was absorption into Hugging Face. The public-compute literature has focused on the gap in training and inference compute. It has not addressed the gap in inference maintenance funding. The NSF's Pathways to Enable Secure Open-Source Ecosystems (PESOSE) program provides a precedent: it funds organizations responsible for the creation and maintenance of open-source infrastructure \cite{nsfPESOSE}. Extending this model to inference runtimes and conversion tools would treat the software that makes open models runnable as a shared resource. Industry coverage of foundation model releases reports significant infrastructure and maintenance costs associated with running large models at scale, which makes it reasonable to expect that public-facing model releases should include sustained maintenance for the tooling that makes them usable.

These mechanisms address the empirical findings without inventing new regulatory frameworks. A 6(b) report produces data that the FTC can use under existing authority. Model certification adapts an existing tool to a new domain. Sustained maintenance funding follows an existing NSF program model. The \texttt{llama.cpp} case shows that open models cannot remain open in practice if the infrastructure that makes them runnable is left to unstable financing and voluntary contribution. Policy attention to the model itself, as the current federal debate does, risks missing additional layers that determine whether local openness fails or succeeds.

\bibliographystyle{ACM-Reference-Format}
\bibliography{references}

\end{document}